\documentclass[11pt]{article}

\usepackage{aaspp4,psfig,epsf,rotate}

\font\caps=cmcsc10 scaled 1200
\def\etal   {{\sl et~al.\ }}
\def\deg   {{$^\circ$\ }}

\slugcomment{Accepted to the Astronomical Journal (Nov. 1999)}

\lefthead{Venturi \& Taylor}
\righthead{RM structure of 3C\,216}

\begin{document}

\title{~~\\ ~~\\ The Galactic Magnetic Field in the Quasar 3C\,216}

\author{T. Venturi\altaffilmark{1} \& G. B. Taylor\altaffilmark{2}}

\altaffiltext{1}{IRA-CNR, Via Gobetti 101, I-40129 Bologna, Italy; 
tventuri@ira.bo.cnr.it}
\altaffiltext{2}{NRAO, Box 0, Socorro, NM 87801, USA; gtaylor@nrao.edu}

\begin{abstract}

Multifrequency polarimetric observations made with the Very
Long Baseline Array of the quasar 3C\,216 reveal the
presence of Faraday rotation measures (RMs) in excess of 2000 rad m$^{-2}$, 
in the source rest frame,
in the arc of emission located at $\sim 140$ mas from the core. 
Rotation measures in the range $-$300 -- $+$300 rad m$^{-2}$
are detected in the inner 5 mas ($\sim$ 30 parsecs). While 
the rotation measure
near the core can be explained as due to a magnetic field in
the narrow line region, we favor the interpretation for the high 
RM in the arc as due to a ``local'' Faraday screen 
produced in a shock where the jet is deflected by 
the interstellar medium of the host galaxy. 
Our results indicate that a galactic
magnetic field of the order of $\sim 50~\mu$G on a scale $>$ 100 pc
must be present in the ambient medium. 

\end{abstract}



\section{Polarimetric Parsec-Scale Observations of High RM Radio Sources}

While typical extragalactic radio sources have Faraday rotation
measures (RMs) of order 10 rad m$^{-2}$ at arcsecond resolution, there
exists $\sim$30 sources with RMs in excess of 1000 rad m$^{-2}$ (see
Taylor, Barton \& Ge 1995; Carilli \etal 1997; Athreya \etal 1998).
Approximately 40\% of such sources are unresolved at arcsecond
resolution.  The parsec-scale study of the RM distribution in these
radio sources exhibiting an excess of RM is important in order to
understand the scale at which such high RMs originate, and to explain
the origin of the phenomenon itself.  High rotation measures could
originate either within the source itself, given a sufficiently dense
thermal component, or they could be due to an external Faraday screen
of thermal gas and magnetic field.

A large fraction of radio sources with high RM on the arcsecond scale,
i.e. $\sim$ 40\% of the total, are compact steep spectrum radio
sources (CSSs), while the rest are extended radio galaxies.  For a few
extended radio galaxies, such as Cygnus A and Hydra A, both located at
the center of clusters of galaxies, it has been proposed that a highly
magnetized intracluster gas is responsible for the high RM detected
(see respectively Dreher, Carilli \& Perley 1987 and Taylor \& Perley
1993).

Recent polarimetric observations of a few high RM radio sources
carried out at parsec-scale resolution with the Very Long Baseline
Array (VLBA) revealed that the RM distribution differs considerably
from source to source. In the CSS quasar OQ172 (Udomprasert et
al. 1997) the high RM observed on the arcsecond scale originates
within the central parsecs, where values around $\sim$ 40000 rad m$^{-2}$
are observed.  It is therefore likely that the nuclear environment
(i.e. the Narrow Line Region, or NLR) is responsible for the high RM 
observed.  Taylor (1998)
studied 4 quasars with normal to low RMs on the arcsecond scale, and
for three of them found RMs in excess of 1000 rad m$^{-2}$ within a few
mas of the core, while the RM drops abruptly by over an order of
magnitude farther along the parsec-scale jet.  Again, it is likely that the
nuclear environment is responsible for the RM distribution observed.
Similar results were obtained also for 3C\,138 (Cotton \etal 1997). 
Finally Aaron \etal (1998) studied the RM distribution in the CSS
quasar 3C\,309.1, and found small RMs and little depolarization, and
concluded that the NLR environment in this quasar is ``normal'' and
uniform.

The quasar 3C\,216 was found to have high RMs in the single dish
surveys of Tabara \& Inoue (1980) and in VLA observations  made by
Taylor,
Ge \& O'Dea (1995).
In this paper we present and
discuss the parsec-scale RM distribution based on multifrequency
polarimetric VLBA observations carried out at 3.6 cm and 6 cm.

3C\,216 is associated with an optically polarized quasar
located at z=0.668. This object is peculiar, since it exhibits
properties typical both of blazars and of CSS sources.  Detailed
observations carried out at parsec-scale resolution (Barthel \etal
1988, Venturi \etal 1993) revealed superluminal motion in the
inner part of the VLBI jet, which argues in favor of a small
orientation of the VLBI structure to the line of sight. The parsec-
and kiloparsec-scale radio emission are misaligned by $\sim
110^{\circ}$.  High frequency VLA polarimetric observations (Taylor,
Ge \& O'Dea 1995) showed that the source is strongly polarized, and
that the high RM originates mainly in the arcsecond core and in the
northeastern component. In particular their results indicate that the
polarized flux in the nuclear components peaks at $\sim$ 150 mas
southeast of the core.


A Hubble constant H$_0$ = 65 km sec$^{-1}$ Mpc$^{-1}$ and q$_0$ = 0.5
will be used throughout this paper. With this choice of the
cosmological parameters, at the distance of 3C\,216 1 mas corresponds to
6.1 parsec.

\section{Observations and Data Reduction}

The observations, performed on 1996 Nov. 3, were carried out at four
frequencies in the 3.6 and 6 cm bands (see Table 1) using the 10
element VLBA of the NRAO\footnote{The National Radio Astronomy
Observatory is a facility of the National Science Foundation operated
under a cooperative agreement by Associated Universities, Inc.}.  Both
right and left circular polarizations were recorded using 1 bit
sampling for 4 widely spaced IFs of 8 MHz bandwidth each.  The VLBA
correlator produced 16 frequency channels/IF in every 2 second
integration.

Amplitude calibration for each antenna was derived from measurements
of the antenna gain and system temperatures.  The
polarization calibration was performed following a procedure suggested
by Cotton (1993).  Global fringe fitting was performed using the AIPS
task FRING, an implementation of the Schwab \& Cotton (1983)
algorithm.  The fringe fitting was performed using a solution interval
of 4 minutes, and a point source model was assumed. Next, a short
segment of the cross hand data from the strongly polarized calibrator
3C\,345 was fringe fitted in order to determine the right-left delay
difference, and the correction obtained was applied to the rest of the
data.  Once delay and rate solutions were applied, the first and last
channel were omitted, and the data were averaged in frequency over the
remaining 7 MHz.  The data from all sources were edited and averaged
over 30 second intervals using {\caps Difmap} (Shepherd, Pearson \&
Taylor 1994; Shepherd 1997) and then were subsequently self-calibrated within
AIPS.

Finally, the strong, compact calibrator 0552+398 was used to determine
the feed polarizations of the antennas.  We assumed that the VLBA
antennas had good quality feeds with relatively pure polarizations,
which allowed us to use a linearized model to fit the feed
polarizations.  Once these were determined, the solutions were applied
to snapshot observations of 3C\,286 and 3C\,345.  3C\,286 has been
observed to have a polarization angle of 30\deg (Cotton \etal  1997),
so a single R-L phase difference was applied to all 3.6 and 6 cm
frequencies to correct the polarization angles to this value.  As a
check on the absolute polarization angle calibration, 3C\,345 was also
imaged.  Components along the jet of 3C\,345 lie parallel to the jet
axis at 6 cm (Cawthorne \etal 1993, Taylor 1998).  
Assuming that this remains the
case over time, we found that the jet component was within 5\deg of
the expected value at 6 cm.  Of prime importance to this experiment
are the relative angles between frequencies, and these were
preserved by applying a constant correction to all frequencies within
each band.

\section{Results}

\subsection{Total intensity images and parsec-scale morphology}

Our total intensity VLBA images were obtained using all four frequencies in
both the 3.6 and 6 cm bands. This provided a very good $u,v$ coverage 
and allowed high sensitivity imaging.
Natural weighted total intensity images at 6 cm and 3.6 cm are shown
in Fig.~1.
Details about the images are given in Table 2.

In both bands the parsec-scale structure of 3C\,216 extends $\sim$ 140
mas in p.a. $\sim 150^{\circ}$, south-east of the northernmost, most
compact component, which we assume to be the core, as our data demonstrate
(see below). The images in both bands show that the jet within the
first 40 mas from the core is not straight, but wiggles with an
amplitude of $\sim 8^{\circ}$.  The jet has a knotty structure, with
two brightness peaks at $\sim$ 7 mas and $\sim 15$ mas from the core
dominating the jet emission.  Our image also indicates that the
parsec-scale jet is not continuous. It remains collimated and readily
visible, though diffuse, out to $\sim$ 40 mas from the core, then the
surface brightness drops, the jet becomes invisible, then
``reappears'' at $\sim 70$ mas from the core.

At $\sim$ 140 mas from the core the radio jet bends sharply by 
an apparent angle of $\sim 90^{\circ}$, 
to form the most striking feature of our total intensity
images, i.e. the extended emission located at $\sim$ 140 mas from the
core, which we will refer to as the ``arc''. The total extent of the
parsec-scale morphology corresponds to a projected linear distance of
$\sim$ 0.85 kpc for our choice of cosmological parameters.  The
extension of the parsec-scale radio emission in 3C\,216 out to $\sim
140$ mas from the core was first revealed by EVN 50 cm and 18 cm
observations (Fejes, Porcas \& Akujor 1992), and its detailed
morphology was first imaged at 18 cm in a global VLBI experiment
(Akujor, Porcas \& Fejes 1996). Our images agree even in the details
of the earlier 18 cm image.
The $u,v$ coverage and resolution of our data do not allow us to image
the sub-kiloparsec radio emission beyond the ``arc''. From the
comparison of our images and those available in the literature,
with resolutions ranging from a fraction of an arcsecond 
(Taylor, Ge \& O'Dea 1995), to 30 milliarscecond (Fejes, Porcas
\& Akujor 1992), to the milliarcsecond resolution of the images
presented in this paper, we suggest that the radio emission from 
the sub-kiloparsec scale jet and arc continues to the south-west
and feeds the south-western knot visible on the arcsecond scale
(see images in Fejes, Porcas \& Akujor 1992 and in Taylor \etal 1995)
This suggestion is consistent with the polarization properties
of 3C\,216 on the arcsecond scale. Taylor \etal
found that the bright knot 
located 1 arcsecond south-west of the core is less depolarized than the
north-eastern one, and it is therefore expected to be closer to
the observer.

%

We computed the point-to-point spectral index of the parsec-scale
components in 3C\,216 between 3.6 cm and 6 cm, $\alpha^{3.6}_6$
($S \propto \nu^{-\alpha}$).  To this aim we used images made with the
same $u,v$ coverage and convolved with the same restoring beam. In
Fig.~2 the 6 cm total intensity contours are superposed on
the grey scale image of the spectral index in the inner
30 parsec. As is clear from the figure, the
northernmost component of the parsec-scale morphology has the most
inverted spectrum, with $\alpha^{3.6}_6 = -0.6$.  The spectral
index steepens along the inner jet. Peaks in the spectral index
distribution, i.e. flattening of the spectrum, coincide with
brightness peaks along the jet.

The arc of emission at $\sim$140 mas from the core has a considerably
steeper spectrum in our two bands. Comparison of the total flux in the
arc derived from our images leads to a value of the spectral index
$\alpha^{3.6}_6 \sim 1$.  The point-to-point distribution (not shown
here) is irregular, and regions of flatter spectrum are located in
between the arc brightness peaks. We compared our flux measurements to
the 18 cm image published in Akujor, Porcas \& Fejes (1996) after
tapering our images to match the published 18 cm image, and we derived
the spectrum of the arc using the information in these three bands.
The result is plotted it in Fig.~3. The spectrum is almost flat
between 18 cm and 6 cm, with $\alpha^{6}_{18} \sim 0.2$, then steepens
to a value of $\alpha^{3.6}_6 \sim 1$.

\subsection{Polarization Images and Rotation Measure structure}

The fractional polarization and polarization percentage
values are reported in Table 3 and
the image of the polarized flux (grey scale) with
superposed total flux density contours are given in Figures 4 and 5 for the
6 cm image (inner jet and arc respectively) and in Fig.~6 for
the inner region in the 3.6 cm image. The total intensity and
polarized intensity images used to derive the
values in Table 3 were made using a single IF at both frequencies
at matching resolution.
To this aim we made a uniformly weighted image at 6 cm and a naturally
weighted and tapered image at 3.6 cm.

As is clear from Table 3 and from the images, the polarized intensity
of the parsec-scale radio emission of 3C\,216 varies considerably
along the structure.  The core region is complex.  The morphology of
the polarized emission is very similar in both bands, and consists of
three components, one coincident with the radio core, and the other
two, labeled K1 and K2 in Fig.~6, along the inner jet, located
respectively at 4 mas and 6 mas from the core.  The core is very
weakly polarized, while components K1 and K2 are polarized at a level
of 6\% and 20\% in both bands. The core appears to depolarize by a
factor of $\sim$2 between 3.6 and 6 cm.  The arc of emission is
strongly polarized, i.e.  $\sim$ 30\% level, in both bands.  Our
results indicate that no depolarization is found between 6 cm and 3.6
cm in the inner jet and in the arc of 3C\,216.

The orientation of the projected magnetic field (already corrected for
the RM) is shown in Figures 7 and 8 for the core and inner jet and for
the arc respectively.  The magnetic field is parallel to the jet axis
near the core and in knot K2, while it is oriented perpendicular to
the jet axis in knot K1. The orientation of the magnetic field also
shows some structure in the arc (see Fig.~8) and tends to remain
parallel to the outer edge.  The high polarization percentage in the
arc and the orientation of the magnetic field vector, coupled with its
morphology, are suggestive of a strong interaction with an external
medium.

We derived the RM in 3C\,216 between 6 cm and 3.6 cm. Our result is
shown in Fig.~9, where total intensity contours at 6 cm are
superposed on a false color map of the rotation measure. The RM values
derived over the whole region imaged 
are in the range $\sim -$100 to $\sim$ 800 rad m$^{-2}$. Since the
rotation measure in the rest frame of the source depends on the
redshift through the simple relations RM$_{\rm intr}$ = RM$_{\rm obs}(1+z)^2$,
in the case of 3C\,216 the intrinsic rotation measure is in the range
$\sim -$280 to $\sim$ 2200 rad m$^{-2}$.

The observed RM in the inner jet region is low, i.e. from $\sim -$100 to 
$\sim$ 100 rad m$^{-2}$. It
flips from $105 \pm 8$ rad m$^{-2}$ to $\sim -99 \pm 12$ rad m$^{-2}$
from the core to the polarized knot K1 at $\sim$ 4 mas
(see Fig.~6), then changes
back to $\sim 100$ rad m$^{-2}$ in knot K2. 
The fits to the polarization angle, derived 
using all four frequencies in the two bands, are shown in Fig.~10.
The change in sign of the RM requires a magnetic field reversal in
the inner jet region.
The largest observed RM values, ranging from $\sim$ 500 to $\sim$ 800 
rad m$^{-2}$, are found
in the arc, in agreement with the RM distribution on the arcsecond
scale (Taylor \etal 1995).
Fits of the RM to the observed polarization angle versus
the square of the wavelength for the arc are given in Fig.~11.
In both regions the variation of the position angle is well fit 
by a $\lambda^2$ law.
We note that the galactic contribution to the measured RM is
in the range 0 -- 10  rad m$^{-2}$ (Simard-Normandin, Kronberg \& Button 1981),
and therefore negligible.

\section{Discussion}

Our study on the parsec-scale polarimetry and rotation measure
structure in the compact steep spectrum quasar 3C\,216 shows that the
arc of emission located at $\sim 140$ mas from the core exhibits the
most extreme properties, with an intrinsic rotation measure in excess
of 2000 rad m$^{-2}$. The rotation measure in the core region, on the
other hand, is an order of magnitude smaller and ranges from
$\sim -300$ to $\sim$ 300 rad m$^{-2}$ with at least two sign
reversals.  The polarization percentage is also much
higher in the arc than in the core region,
independent of our two observing bands.

Since no depolarization is found in the jet or in the arc
between 3.6 cm and 6 cm, even while the polarization angle changes by 
more than 100$^{\circ}$ in the arc,
the most likely mechanism for the high rotation measure observed
in 3C\,216 is the presence of an external screen.
Internal Faraday rotation is ruled out, as it would
imply also considerable depolarization with frequency (Burn 1966),
which is not observed.  The lack of depolarization also argues against
more complex situations, such as for example an external screen with
magnetic field tangled on scales less than the beamsize
(see Feretti \etal 1995 for a brief review on
the subject). The situation in the core itself is less clear, since
there is very low fractional polarization, and evidence of
depolarization.

If our assumption is correct, then the intrinsic RM is
related to the electron density $n_e$ (cm$^{-3}$), to the component of the 
magnetic field parallel to the line of sight $B_{||}$ ($\mu$G) and to the 
depth of the screen $d$ (kpc) through the formula

\centerline{$<RM>=812 n_e B_{||} d$ rad m$^{-2}$.}

\subsection{The core and the inner jet}

The screen responsible for the rotation measure
distribution in the inner jet region of 3C\,216 is most likely the
NLR, given that it is all contained within a distance of $\sim 5$ mas
from the core, a projected distance of $\sim 30$ parsec.
From the intrinsic RM obtained in our observations, i.e.  $|RM_{\rm
intr}| \sim 300$ rad m$^{-2}$, and assuming a depth of 10 pc,
we derive $n_e B_{||} \sim 35$ cm$^{-3}\mu$G.  With this constraint
we can derive an estimate of the component of the magnetic field
parallel to the line of sight for the narrow line region of the order
of 10$^{-2} \mu$G, as suggested by Taylor (1998), only if we assume an
electron density of the order of 3$\times 10^3$ cm$^{-3}$, a lower limit for
the density of the NLR. The sign reversal of the RM in the inner jet region,
however, suggests that the magnetic field could be tangled on the
assumed 10 pc scale, which implies stronger fields than estimated
here.

These properties are different from the results reported in
Taylor (1998) for a sample of quasars, where Faraday rotation measures
in excess of 1000 rad m$^{-2}$ are found within
the first 10 -- 20 parsec from the quasar core, and 
drop considerably further out.  Our results are unexpected, considering
that the inner region of 3C\,216 shares the same global properties of
the sources in Taylor sample, such as for example superluminal motion
and low fractional polarization. These similar properties would lead
to the conclusion that we are observing the same type
of sources, all seen at a small viewing angle.

One explanation for this inconsistency is that the resolution of our
3C\,216 images is too low to allow us to separate the core from the
jet beginning, and therefore we lack information on the properties of
the true core. The depolarization of the core between 3.6 and 6 cm
could also be indicative of large RM gradients within the beam.  The
RM distributions of Taylor (1998) were derived between 8 GHz and 15
GHz, consequently with higher resolution.  To test this hypothesis we
made high resolution 3.6 cm images of the polarized intensity and
rotation measure distribution, to see if there is an indication of an
increase in the rotation measure in the northern extreme of the
source. A RM of $\sim 1500 \pm 500$ is found, and we suggest that the
core may be located at the extreme end of the most compact component
in our images. Observations at higher resolution and higher frequency
are needed to investigate if the Faraday rotation in the core of
3C\,216 follows the behavior which now seems common in quasar cores.

\subsection{The arc}

The external screen responsible for the high Faraday rotation in the arc
must be located quite far from the central engine. As
stated in Venturi \etal (1993), the superluminal motion in the inner
6 pc argues in favor of a small angle of the parsec-scale jet to the
line of sight, i.e.  $\theta \le 20^{\circ}$. The global morphology on
the parsec-scale does not indicate misalignments or bending after the
first few parsecs, so it seems reasonable to assume that the arc
where the jet bends is still seen at the same angle. If our
assumptions are correct, then the true distance of the arc from the
core of 3C\,216 is $\sim$ 2.5 kpc. Such a distance could still be within
the NLR, though it would be a somewhat extreme situation (Netzer
1991, Fanti \etal 1995), or it could be 
within the interstellar medium, possibly inside a shell surrounding 
the radio emission.
Another possibility is that the arc is the result of a shock between
the radio emission and the external medium. Under this hypothesis the 
Faraday screen would be ``local'', in a thin sheet
of shocked gas around the radio emission. In the following we will
consider both frameworks and discuss their implications.

\subsubsection{Galactic Faraday screen}

Under the assumption that the Faraday screen has a depth of
the order of a kiloparsec, for a value $<RM> = 2000$ rad m$^{-2}$ we
obtain $n_e B_{||} \sim$ 2.5 cm$^{-3}\mu$G. 
The interstellar gas density at this distance from the nucleus 
in high redshift objects is not known. Studies carried out at
X-ray energies on low redshift galaxies indicate the existence
of hot coronae around galaxies, with densities of the
order of 10$^{-1}$ cm$^{-3}$ and core radii of the order of
few kiloparsecs (Forman \etal 1985, White \& Sarazin 1988, see also
Fanti \etal 1995 for a brief review on this topic).
If we assume that the same condition
holds also in distant ellipticals and quasars, such density would lead
to a magnetic field strength in the interstellar medium
(for the component parallel to the line of sight) of $\sim $ 25
$\mu$G. Assuming that the field is ordered (as the positive rotation
measure throughout the arc suggests), but not entirely along the line
of sight, then a more accurate estimate for the magnetic field
strength is $\sim 25 \times \sqrt 3 \sim 43~ \mu$G.
A very similar value was found for the magnetic field in the galactic
gas surrounding M87 (Owen, Eilek \& Keel 1990).

\subsubsection{Bow shock Faraday screen}

The sub-kiloparsec scale morphology of 3C\,216, characterized by a
very faint jet which flares into the southern arc
and bends westward here, suggests
that the arc could be the working surface of a sub-kiloparsec jet in
the galactic medium, and that the magnetized screen responsible for the Faraday
rotation would be a thin layer of compressed galactic medium.
A similar bow shock has been reported on somewhat larger
scale for Cygnus A (Carilli, Perley \& Dreher 1988).
The interaction between the jet and the compressed ambient medium 
could be responsible for the jet deflection.
We note here that with a viewing angle of $\sim 20^{\circ}$,
as we have assumed throughout the paper, the apparent bend of 
$\sim 90^{\circ}$ corresponds to an intrinsic bend of $\sim 75^{\circ}$.

Under this hypothesis the smoothness of the RM distribution 
suggests a plausible depth of 100 pc for such a screen.
This leads to an estimate of the product $n_e B_{||} \sim 25$ cm$^{-3}\mu$G.
If the average  density out of the shocked region is 0.1 cm$^{-3}$ 
(as justified in \S 4.2.1) and with a compression
factor of 4 for a strong adiabatic shock,
the electron density just behind the shock is 0.4 cm$^{-3}$,
thus leading to an estimate of $B_{||} \sim 60~ \mu$G in the shocked
region. The galactic magnetic field outside the shock is expected 
to be 15$\sqrt 3 \sim 26 ~\mu$G.
Given the uncertainties in the assumptions
made we can say that these two estimates are in agreement.

The external density could be much higher than
assumed here, if the jet is deflected by interaction with a dense 
molecular cloud in the narrow line region. For this reason the value we 
derive for the galactic magnetic field in 3C\,216 under this assumption
should be considered an upper limit. 
However, three-dimensional hydrodynamic simulations of deflected 
cosmic jets (de Young 1991) show that jet-cloud interactions
are not a stable mechanism for jet deflection, even in cases
with an optimal geometry and choice of parameters for the deflection. 
This mechanism considerably reduces the jet speed on a short time scale
($t < 10^7$ yr), and leads to the cloud eroding on timescales $\ge 10^7$
yr. Furthermore, de Young (1991) argues that such interaction would not 
result in a coherent observable bent jet.
This is not in agreement with the radio emission in 3C\,216, 
which indicates that the sub-kiloparsec scale jet is smoothly bending
and leading into the arcsecond scale emission in the
source, as we have pointed out in \S 3.1.


Using equipartition arguments, with the standard assumptions
(i.e. filling factor $\phi$ = 1, k=1 and integrating between 10 MHz
and 100 GHz) and using the spectral index $\alpha_{6}^{3.6}$ = 1
derived in Section 3.1, we derive an equipartition magnetic field
in the arc $B_{\rm eq}$ = 1.3 mG and an internal non thermal pressure
$P_{\rm int} = 2.6 \times 10^{-7}$ dyne cm$^{-2}$. 

\section{Conclusions}

Detailed polarimetric studies of compact sources 
on the parsec-scale have demonstrated that there are a variety
of mechanisms that can add RMs in excess of 1000 rad m$^{-2}$.
These range from the narrow line region on scales of a few parsecs
(e.g. OQ\,172 -- Udomprasert \etal 1997; and probably in the core
of 3C\,216 as well); to galactic magnetic fields organized on 
scales of 100s to 1000s of parsecs (see \S 4.2); to cluster magnetic
fields organized on scales of 10s of kiloparsecs (e.g. 3C\,295 -- 
Perley \& Taylor 1995).  Clearly measuring an integrated RM by
itself is not sufficient to determine the nature of the Faraday screen.

Given the parsec-scale morphology of 3C\,216 we favor the bow shock model to 
explain the change in the jet direction and the presence of 
high Faraday rotation measures.  In this picture, magnetic 
fields of $\sim 50$ $\mu$G organized on scales of $>$100 pc are 
required.

\acknowledgments

The authors wish to thank D. Dallacasa for the many discussions
while this paper was in progress. R. Fanti and C. P. O'Dea
are warmly acknowledged for helpful comments on the manuscript.

This research has made use
of the NASA/IPAC Extragalactic Database (NED) which is operated by the
Jet Propulsion Laboratory, Caltech, under contract with NASA.  
 
\def\dg{$^{\circ}$}
\begin{center}

\clearpage 
TABLE 1 \\
\smallskip
VLBA O{\sc bservational} P{\sc arameters}
\smallskip
 
\begin{tabular}{l r r r r r r r r}
\hline
\hline
Source & Frequency & BW & $\Delta$t & Scan & Time \\
(1) & (2) & (3) & (4) & (5) & (6) \\
\hline
\noalign{\vskip2pt}
3C\,216 & 8.114, 8.309, 8.506, 8.594 & 7 & 2 & 16 & 208 \\
        & 4.616, 4.654, 4.854, 5.096 & 7 & 2 & 16 & 224 \\
0552+398 & 8.114, 8.309, 8.506, 8.594 & 7 & 2 & 5 & 63 \\
        & 4.616, 4.654, 4.854, 5.096 & 7 & 2 & 5 & 67 \\
3C\,286 & 8.114, 8.309, 8.506, 8.594 & 7 & 2 & 6 & 6 \\
        & 4.616, 4.654, 4.854, 5.096 & 7 & 2 & 5 & 5 \\
3C\,345 & 8.114, 8.309, 8.506, 8.594 & 7 & 2 & 6 & 12 \\
        & 4.616, 4.654, 4.854, 5.096 & 7 & 2 & 6 & 12 \\
\hline
\end{tabular}
\end{center}
\smallskip
\begin{center}
{\sc Notes to Table 1}
\end{center}
Col.(1).---Source name.
Col.(2).---Observing frequency in GHz.
Col.(3).---Total spanned bandwidth in MHz.
Col.(4).---Integration time output from correlator in seconds.
Col.(5).---Approximate scan length in minutes.
Col.(6).---Total integration time on source in minutes.
\bigskip
 
\begin{center}
TABLE 2 \\
\smallskip
P{\sc arameters} from the F{\sc inal} T{\sc otal} I{\sc ntensity} I{\sc mages}
\smallskip
 
\begin{tabular}{l c l r r r}
\hline
\hline
Frequency & FWHM & rms & $S_{\rm core}$ & $S_{\rm jet}$ & $S_{\rm arc}$ \\
(1) & (2) & (3) & (4) & (5) & (6) \\
\hline
\noalign{\vskip2pt}
4.805  & 2.65$\times$1.75, 0$^{\circ}$ & 0.07 & 513.5 & 86.1 & 83.5 \\
8.381  & 2.65$\times$1.75, 0$^{\circ}$ & 0.10 & 707.6 & 73.2  & 32.9 \\
\hline
\end{tabular}
\end{center}
\smallskip
\begin{center}
{\sc Notes to Table 2}
\end{center}
Col.(1).---Observing Frequency.
Col.(2).---Restoring beam of the images presented in Figs. 1a and 1b, in mas.
Col.(3).---The rms noise level in the final image given in mJy/beam.
Col.(4).---Core flux density (mJy).
Col.(5).---Flux density of the inner 30 mas of the parsec-scale jet (mJy).
Col.(6).---Flux density of the bent component located at $\sim$ 140 mas (mJy).
\bigskip
 
\begin{center}
\clearpage 
TABLE 3 \\
\smallskip
P{\sc arameters} from the F{\sc inal} P{\sc olarized} I{\sc ntensity} 
I{\sc mages}
\smallskip
 
\begin{tabular}{l c l r r r r r r r r}
\hline
\hline
Frequency & FWHM & rms & $P_{\rm core}$ & $m_{\rm core}$ & $P_{\rm K1}$ 
& $m_{\rm K1}$ & $P_{\rm K2}$ & $m_{\rm K2}$ &
$P_{\rm arc}$ & $m_{\rm arc}$ \\
(1) & (2) & (3) & (4) & (5) & (6) & (7) & (8) & (9) & (10) & (11) \\
\hline
\noalign{\vskip2pt}
4.854  & 2.65$\times$1.75, 0$^{\circ}$ & 0.14 & 1.5  & 0.3 & 1.8 & 6
& 2.9 & 21 & 14.4 & 29 \\
8.309  & 2.65$\times$1.75, 0$^{\circ}$ & 0.18 & 3.9  & 0.6 & 1.7 & 6
& 2.2 & 20 & 8.6 & 30 \\
\hline
\end{tabular}
\end{center}
\smallskip
\begin{center}
{\sc Notes to Table 3}
\end{center}
Col.(1).---Observing Frequency.
Col.(2).---Restoring beam of the images presented in Figs. 3a, 3b, 4a and 4b, 
in mas.
Col.(3).---The rms noise in the final polarized image given in mJy/beam.
Col.(4).---Polarized flux density of the core (mJy).
Col.(5).---Fractional polarization in \% for the core.
Col.(6).---Polarized flux density of knot K1 (mJy).
Col.(7).---Fractional polarization in \% for knot K1.
Col.(8).---Polarized flux density of knot K2 (mJy).
Col.(9).---Fractional polarization in \% for knot K2.
Col.(10).---Polarized flux density of the arc (mJy).
Col.(11).---Fractional polarization in \% for the arc.

\clearpage 


\begin{figure}
\centerline{\psfig{figure=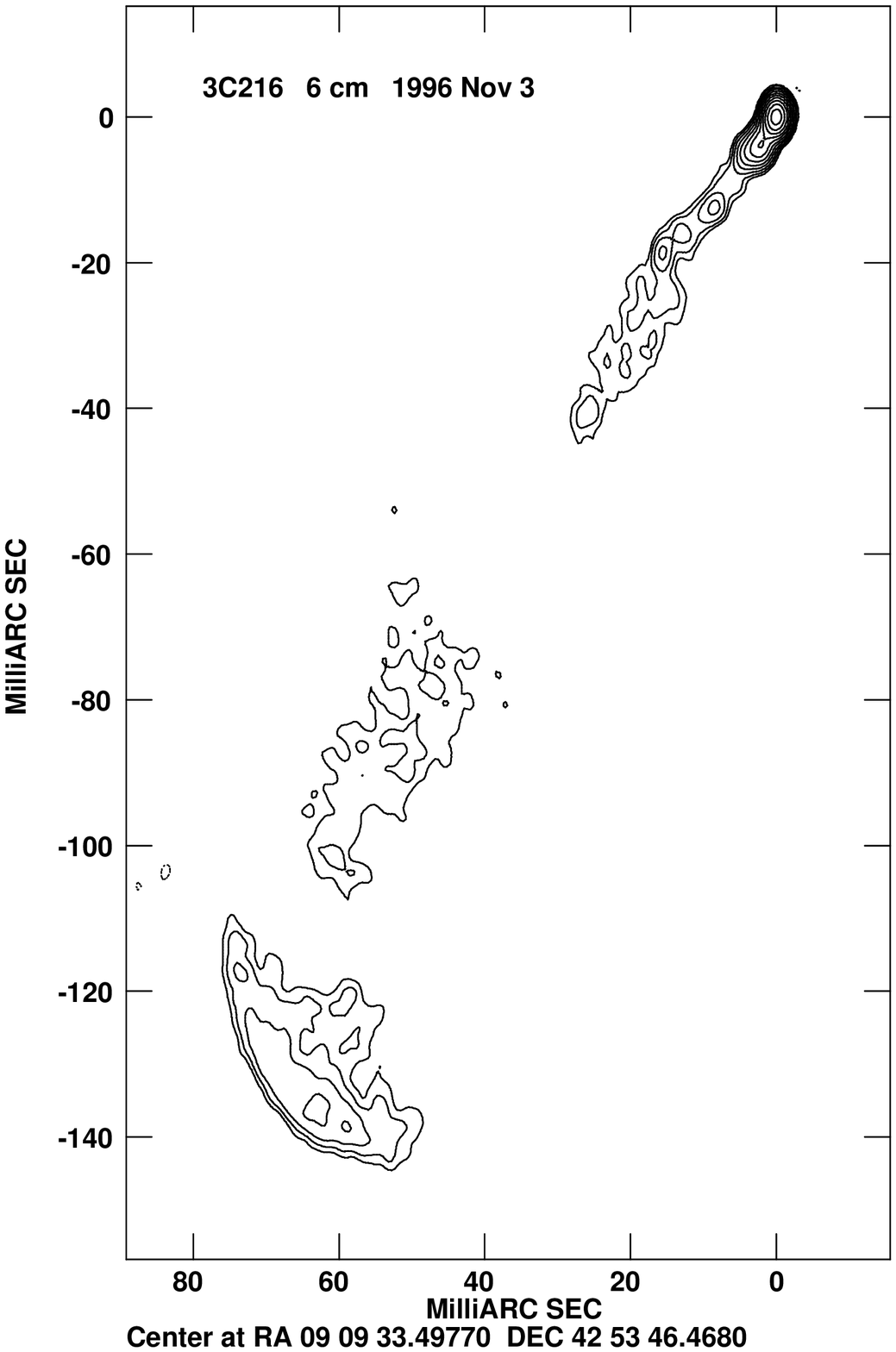,width=8.0truecm},\psfig{figure=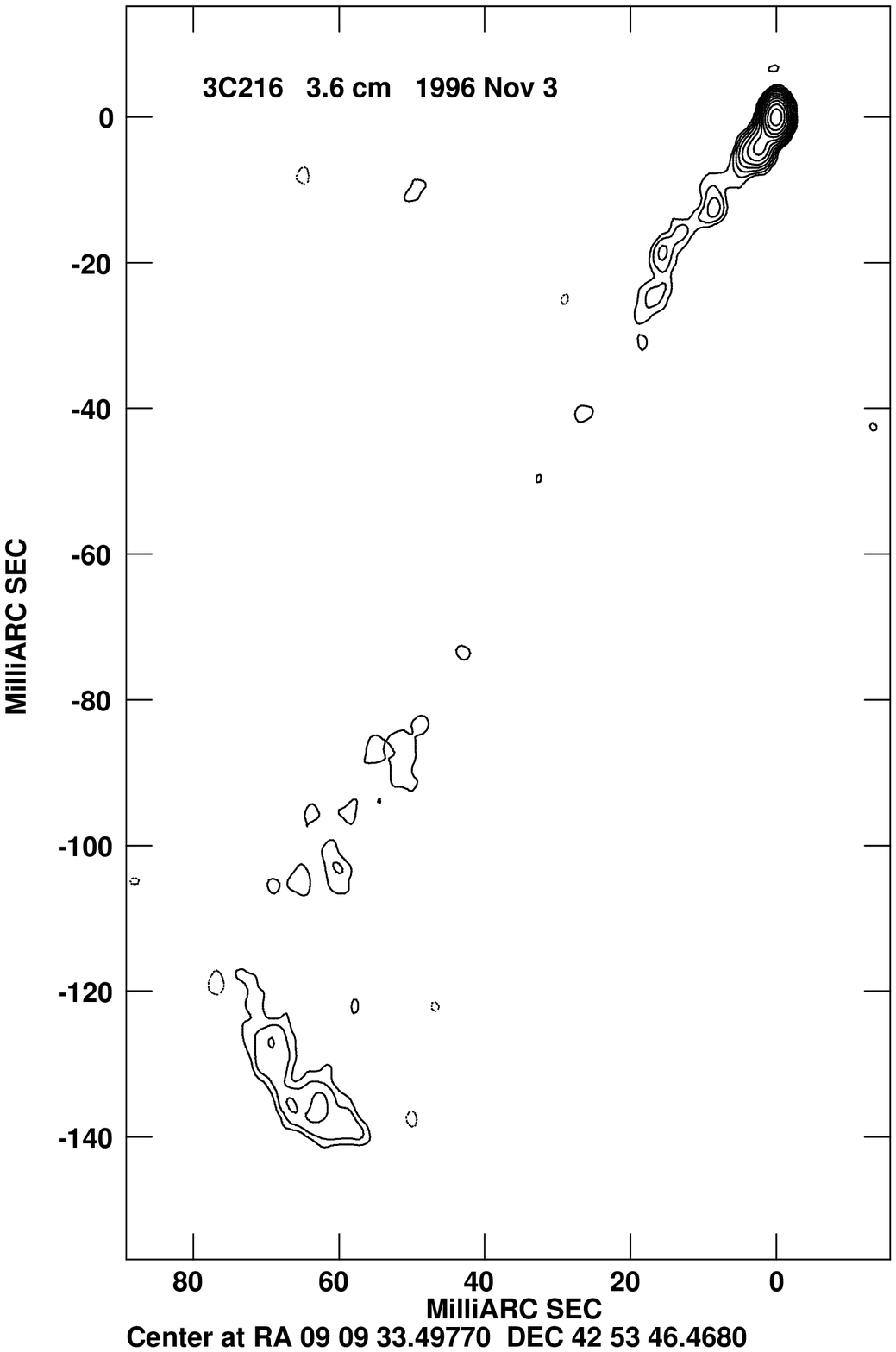,width=8.0truecm}}
\figcaption{Full resolution images of 3C\,216 at 6 cm (left) and 3.6 cm
(right). The restoring beam is 2.65 $\times$ 1.75 mas in position
angle 0$^{\circ}$. Contour levels start at 0.4 (6 cm) and 0.5 mJy (3.6
cm) and increase by factors of 2 up to 410 and 512 mJy/beam at 6 and
3.6 cm respectively.
\label{clnmaps}}
\end{figure}
\clearpage

\begin{figure}
\centerline{\psfig{figure=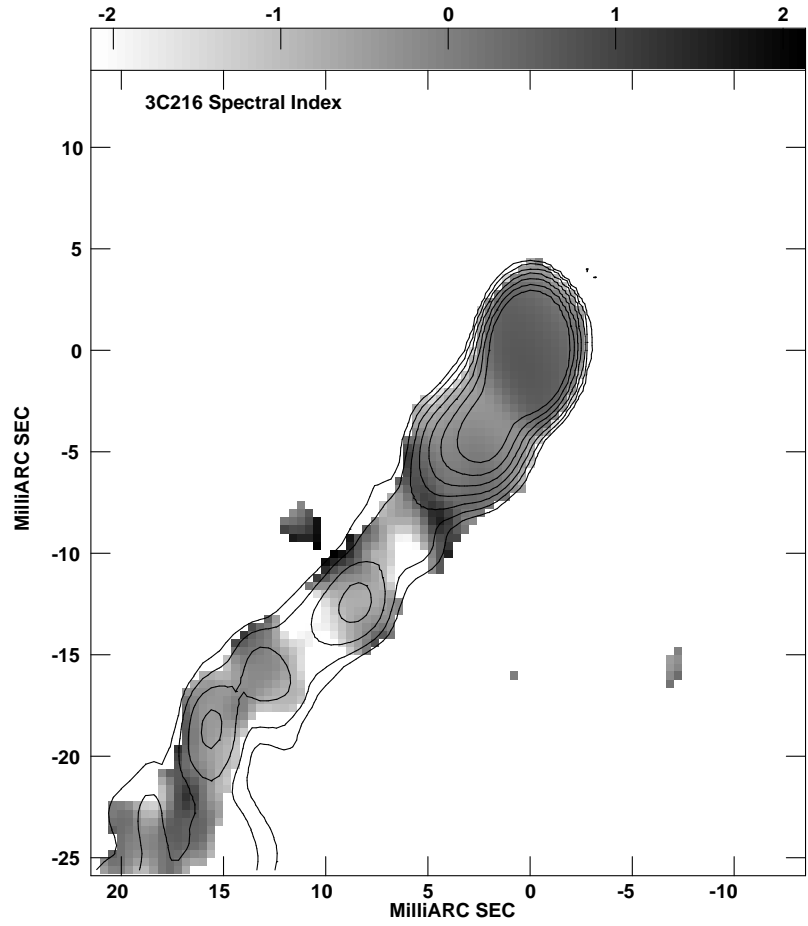,width=10.8truecm}}
\figcaption{Image of the spectral index distribution between 3.6 cm and 6
cm, with superposed flux density contours at 6 cm for the region
within the inner 30 mas. Contour levels are at $-$0.4, 0.4, 0.75, 1, 
3, 5, 10, 50, 100, mJy/beam.
\label{spix}}
\end{figure}

\begin{figure}
\centerline{\psfig{figure=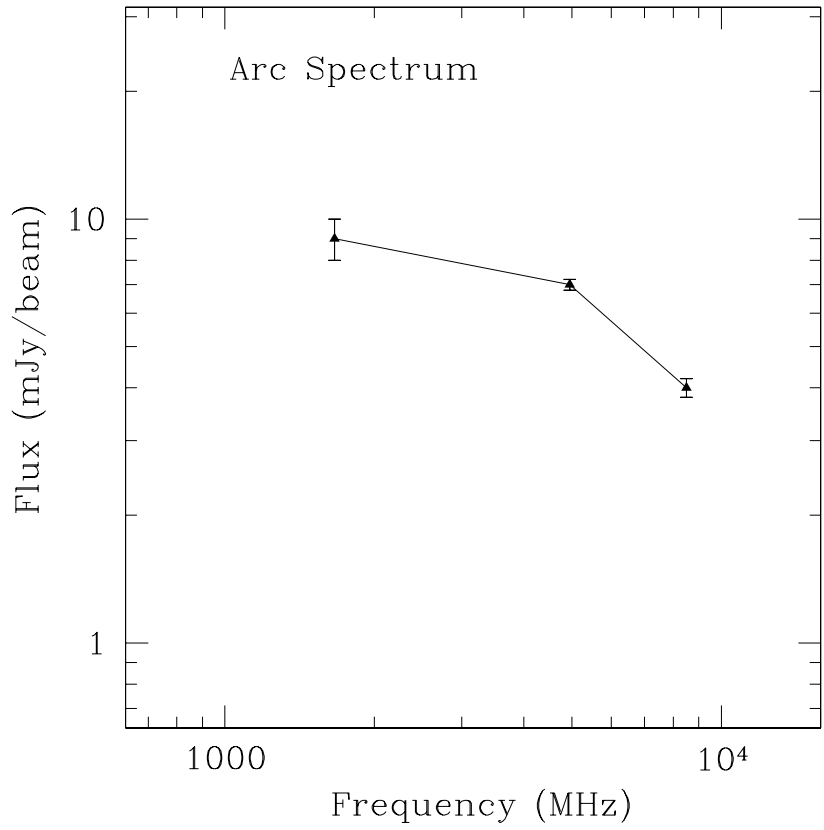,width=10.8truecm}}
\figcaption{Spectrum of the arc in 3C\,216 in the wavelength range 
3.6 cm -- 18 cm.  The 18 cm flux density was estimated from
the image published by Akujor, Porcas \& Fejes (1996).
\label{fig3}}
\end{figure}

\begin{figure}
\centerline{\psfig{figure=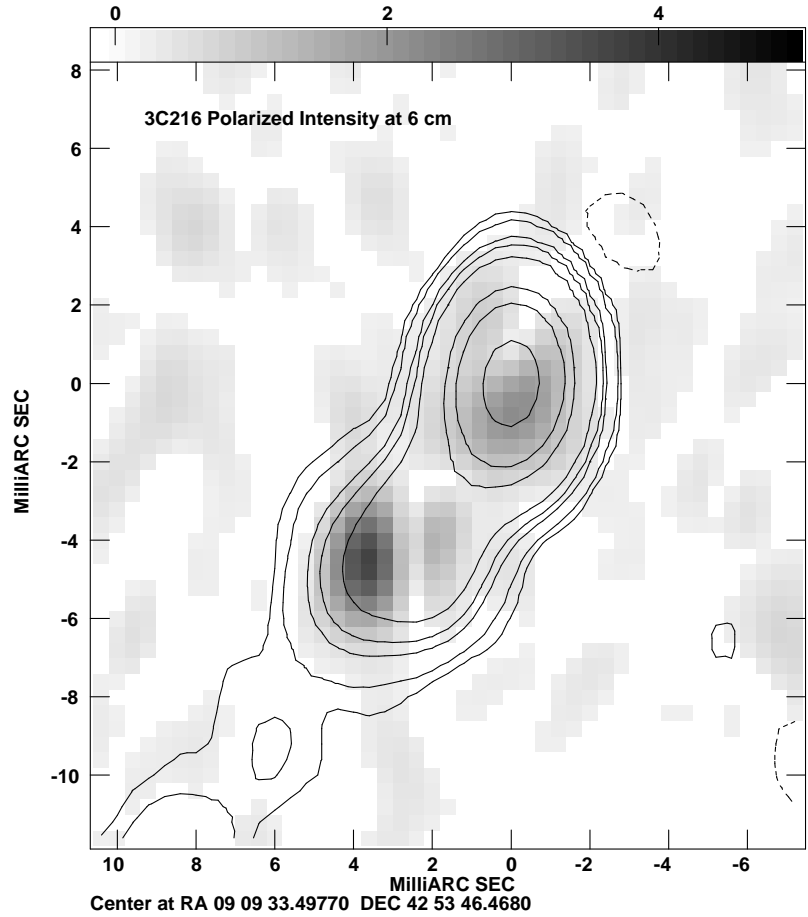,width=10.8truecm}}
\figcaption{Grey scale image of the 6 cm polarized intensity with superposed
flux density contours of the total intensity image for the core region in
3C\,216. Both images are convolved with a beam of 
2.65 $\times$ 1.75 mas in position angle $0^{\circ}$. The grey scale 
ranges from 0 to 5 mJy/beam.  Contour levels are
at $-$0.5, 0.5, 1, 3, 5, 10, 50, 100, 300 mJy/beam.
\label{fig4}}
\end{figure}
\smallskip

\begin{figure}
\centerline{\psfig{figure=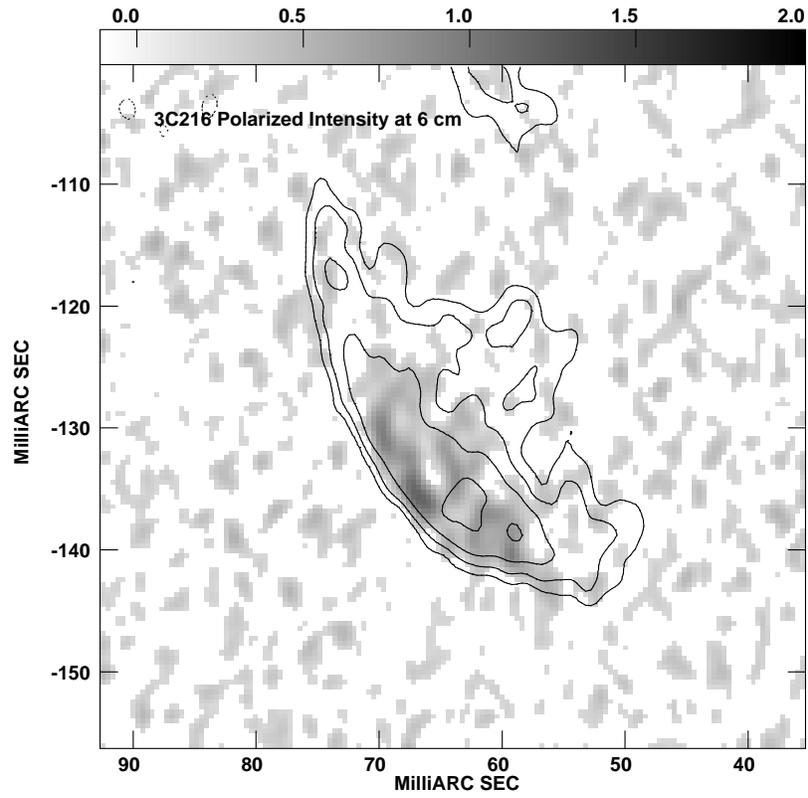,width=10.8truecm}}
\figcaption{Same as Fig. 4 for the arc at $\sim 140$ mas from the core.
Contours are at $-0.3$, 0.3, 6, 1.2, 2.4 mJy/beam. The grey scale ranges
from 0 to 2 mJy/beam.
\label{fig5}}
\end{figure}

\begin{figure}
\centerline{\psfig{figure=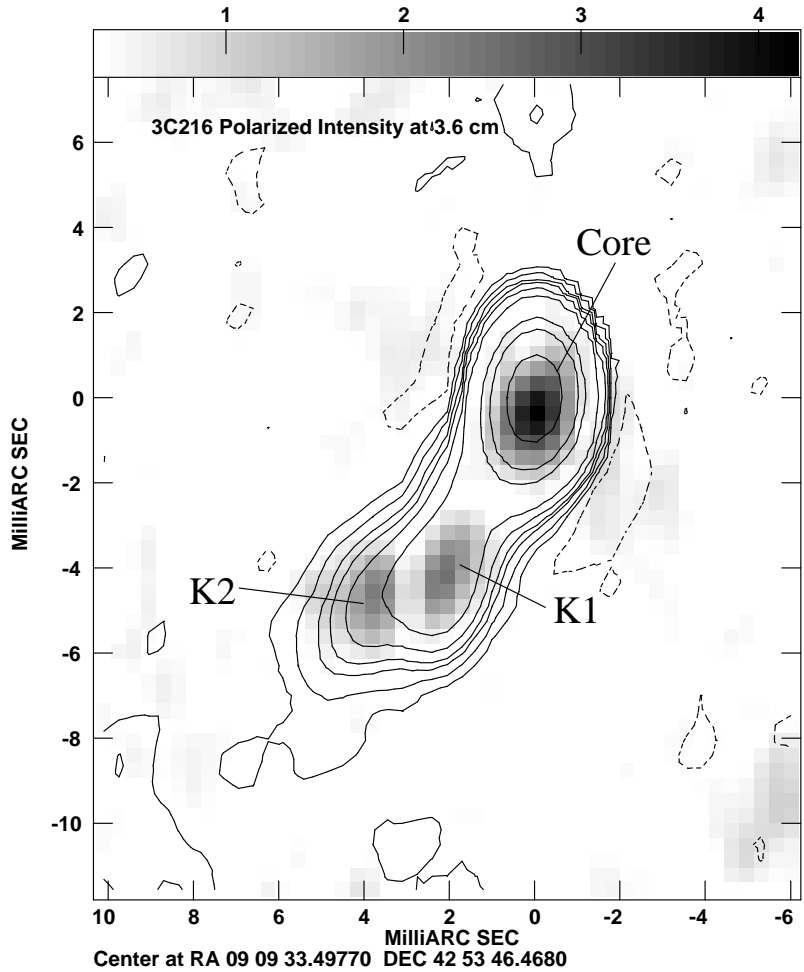,width=10.8truecm}}
\figcaption{Same as Fig. 4 at 3.6 cm. Both images are convolved with a beam of
1.9 $\times$ 1.1 mas in position angle $-0.7^{\circ}$. The grey scale ranges
from 0.3 to 4.1 mJy/beam.  Contour levels are
at $-$0.4, 0.4, 1, 2, 3, 5, 7, 10, 50, 100, 300 mJy/beam.
\label{fig6}}
\end{figure}
\clearpage

\begin{figure}
\centerline{\psfig{figure=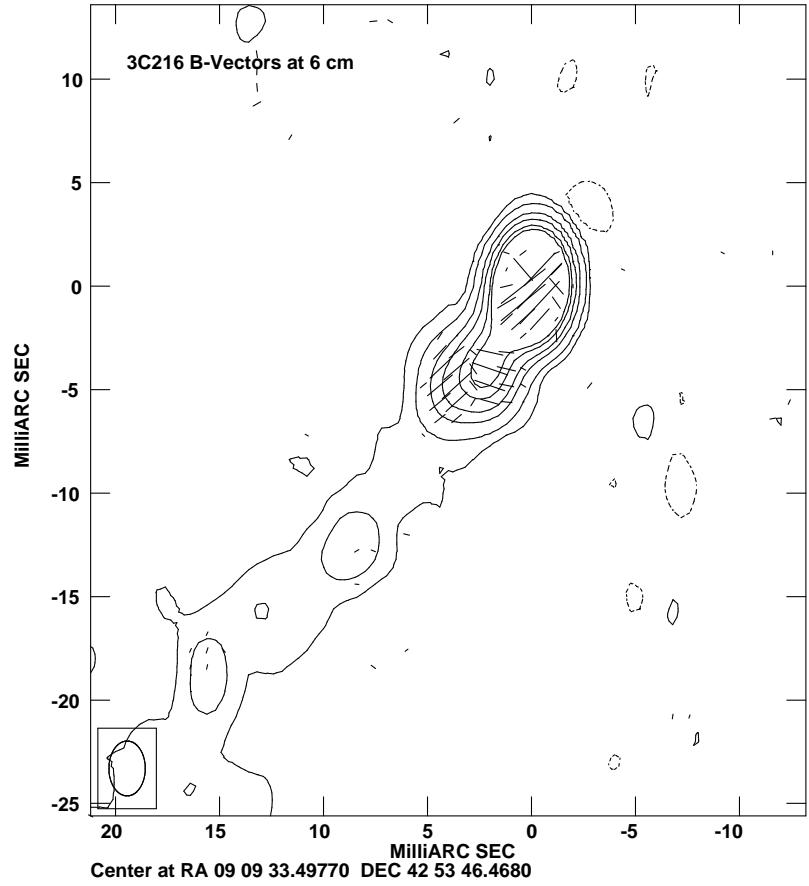,width=10.8truecm}}
\figcaption{The projected magnetic field orientation of the core
and inner jet components at 6 cm after correcting for the RMs shown in 
Fig.~9. The resolution is 2.65$\times$1.75 mas in p.a. 0$^{\circ}$.
Contours of the total intensity image are $-$0.4, 0.4, 1, 1.5, 10, 20,
30 mJy/beam.
\label{fig7}}
\end{figure}
\begin{figure}
\centerline{\psfig{figure=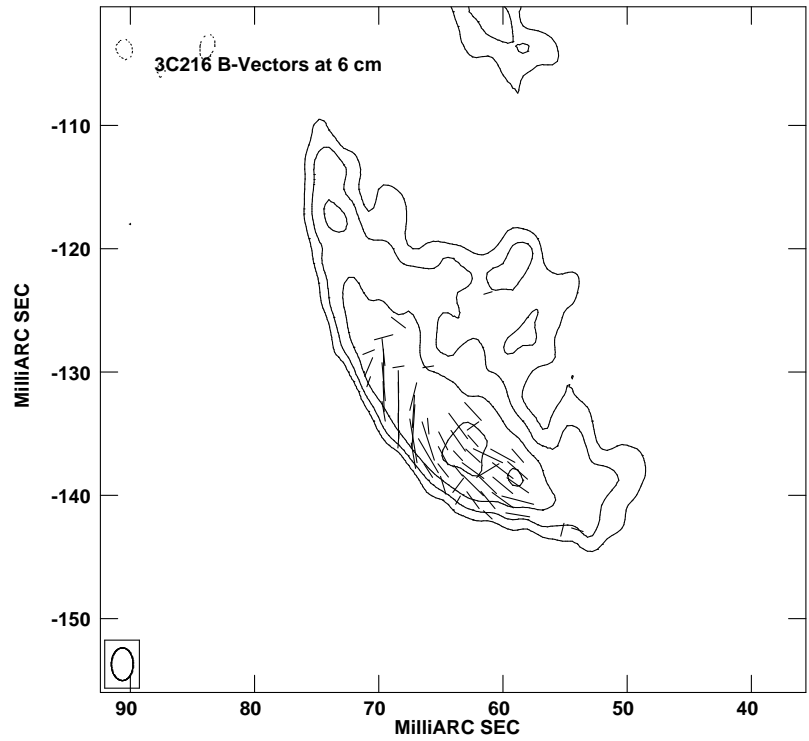,width=10.8truecm}}
\figcaption{Same as Figure 7 for the arc of emission.
Contours of the total intensity image are -0.4, 0.4, 0.75, 1, 1.3 
mJy/beam.
\label{fig8}}
\end{figure}
\clearpage

\begin{figure}
\centerline{\psfig{figure=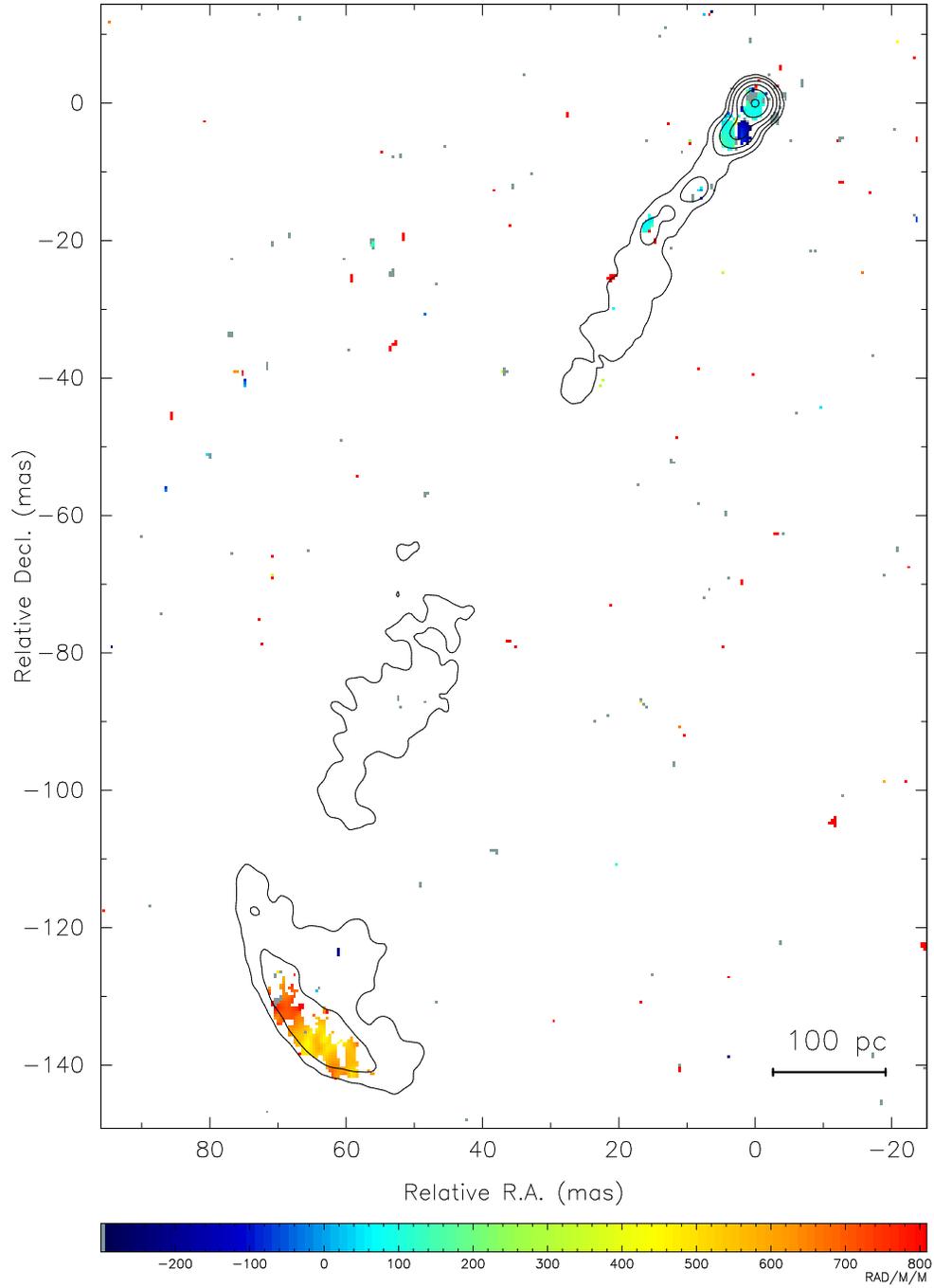,height=18.0truecm}}
\figcaption{Rotation measure image of 3C\,216 with contours of 6 cm total
intensity superposed.
\label{fig9}}
\end{figure}
\clearpage

\begin{figure}
\centerline{\psfig{figure=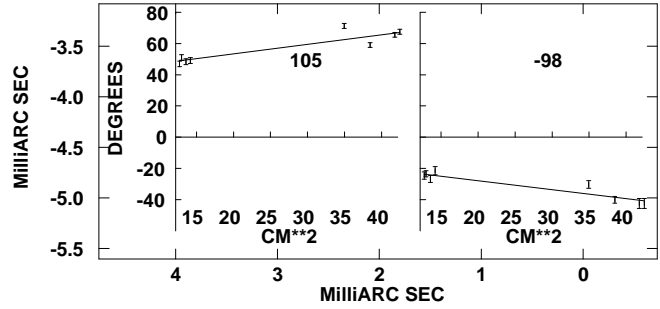,width=8.8truecm}}
\figcaption{Sample fits to the polarization angle versus wavelength
squared in the knots K1 (right) and K2 (left) for the 6 cm and 3.6 cm band.
\label{fig10}}
\end{figure}
\begin{figure}
\centerline{\psfig{figure=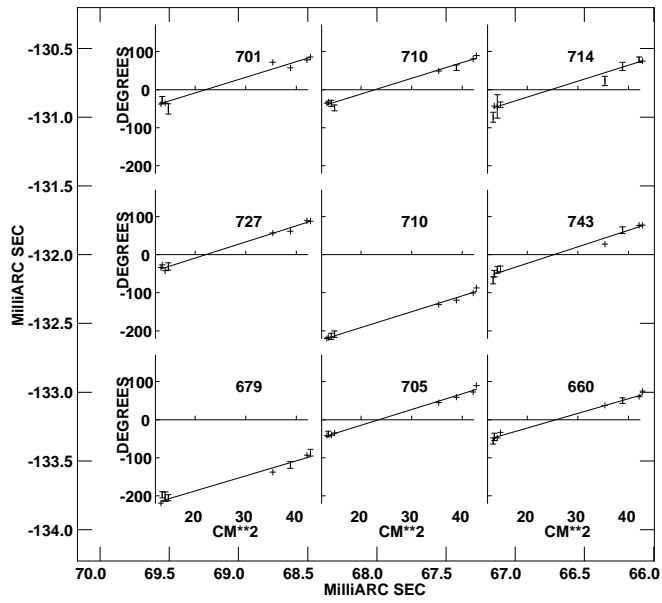,width=8.8truecm}}
\figcaption{Sample fits to the polarization angle versus wavelength
squared in the arc of 3C\,216 for the 3.6 cm and 6 cm bands.
The fits are plotted every third pixel.
\label{fig11}}
\end{figure}
\clearpage

\end{document}